\documentstyle[12pt,epsf]{article}
\newlength{\overeqskip}
\newlength{\undereqskip}
\newcommand{\nc}{\newcommand}
\nc{\be}{\begin{equation}} 
\nc{\ee}{\end{equation}}
\nc{\bea}{\begin{eqnarray}}
\nc{\eea}{\end{eqnarray}}
\nc{\bi}[1]{\bibitem{#1}}
\nc{\lsim}{\mbox{\raisebox{-.6ex}{~$\stackrel{<}{\sim}$~}}}
\nc{\gsim}{\mbox{\raisebox{-.6ex}{~$\stackrel{<}{\sim}$~}}}
\nc{\nn}{\nonumber}

\def\ra{{\rightarrow}}

\def\sl#1{{#1 \!\!\!/ }}
\def\ABS#1{{\vert #1 \vert }}
\def\ave#1{{\langle #1 \rangle}}
\def\nBR#1{{\bigl( #1 \bigr)}}
\def\NBR#1{{\left( #1 \right)}}                 
\def\BBR#1{{\left[ #1 \right]}}                 

\def\DBR#1#2{{\bigl\{#1 \bigr\}
              \bigl\{#2 \bigr\}}}               
\def\DBRb#1#2#3{{\bigl\{#1 \bigr\} #2 
                 \bigl\{#3 \bigr\}}}            

\def\tom{{\tilde \omega}}

\def\calA{{\cal A}}
\def\calC{{\cal C}}
\def\calT{{\cal T}}

\begin{document}
%
%
\begin{titlepage}
\pagestyle{empty}
\baselineskip=21pt
\rightline{NORDITA 2000/11 HE}
\rightline{LPT-ORSAY 00-24}
\rightline{UNIL-IPT/00-03}
\rightline{February 2000}
\vskip .5in
\begin{center}
     {\LARGE {\bf Fermion propagator in a nontrivial} \\
\medskip
                    {\bf     background field}}
\end{center}
\vskip .3in
\begin{center}
   {\large  Michael Joyce} \\
       {\it LPT, Universit\'e Paris-XI, B\^atiment 211,
F-91405 Orsay Cedex, France\\
   E-mail: Michael.Joyce@th.u-psud.fr}  \\
\medskip
   {\large   Kimmo Kainulainen}\\
       {\it NORDITA,
            Blegdamsvej 17, DK-2100, Copenhagen \O , Denmark
    \\E-mail: kainulai@nordita.dk}\\
\medskip
   {\large   Tomislav Prokopec}\\
     {\it Universit\'e de Lausanne, Institut de Physique Th\'eorique,
BSP, CH-1015 Lausanne, Suisse
  \\E-mail: Tomislav.Prokopec@ipt.unil.ch}\\
\end{center}

\vskip 0.3 in

\centerline{ {\bf Abstract} }
\baselineskip=18pt
\vskip 0.5truecm

\noindent
We study the fermion propagator in a spatially varying classical
background field, and show that, contrary to common wisdom, it may get 
nontrivial gradient corrections already at the first order in  
derivative expansion. This occurs whenever the fermion self-energy acquires
a spatially (or temporally) varying pseudoscalar term, a simple example 
of which is given by a complex mass term $\hat m(x)\equiv m_R+i\gamma_5 m_I$.  
Such effective mass terms arise for example in extensions of the 
Standard Model during the electroweak transition, and they are crucial 
in providing the CP-violation necessary for electroweak baryogenesis.  
\\

\end{titlepage}

\baselineskip=20pt

%
%
\section{Introduction}

Often in applications of quantum field theories, it is desirable to treat
a part of the problem semiclassically and compute the quantum corrections
in a controlled expansion around the classical solutions. This applies to
a broad spectrum of problems, including for example the Fermi liquid theory,
as well as the study of the expanding and cooling plasma 
in the heavy ion collisions \cite{Henning, Heinz}.  In 
these, and in many  other cases, it is often sufficient to assume that 
the background is very slowly varying and to treat the problem  
at leading order in gradients.  Such an adiabatic (quasiparticle) 
approximation may make sense if the physical effect of the background 
one is interested in is already present at the trivial order. 
However, there are situations where a semiclassical 
approximation is justified, but the system displays an important 
physical effect only at a nontrivial order in gradients. An important 
example of this occurs at the electroweak transition, where the 
problem is to study particle propagation in the presence of a 
spatially varying, CP-violating Higgs condensate. The CP-violating 
effects of the condensate on the plasma dynamics arise only at the 
first nontrivial order in gradient expansion \cite{CKN,JPT,HN,CJK}. The 
induced perturbations in the particle and antiparticle densities, 
coupled with the weak anomaly, lead to a local baryon production 
mechanism, capable of creating the observed matter-antimatter 
asymmetry of the Universe \cite{SR,CJK}.

Unfortunately, no consistent derivation of the semiclassical quantum 
Boltzman equation for a fermionic field beyond the trivial order in 
gradients exist in the literature. Moreover, there exists a well known 
no-go theorem \cite{Henning}, according to which first nontrivial 
corrections to fermion propagator arise only at {\em second} order in 
derivative expansion. This is in apparent contrast with the classical 
force mechanism (CFM) for baryogenesis \cite{JPT,CJK}, according to 
which the nontrivial 
effects on fermion propagation are felt already at the first order in 
gradients. In this letter we show that the no-go theorem does not hold 
universally, but instead, whenever the fermion self energy contains 
a pseudoscalar term, there are corrections to the propagator already 
at the first order in gradients. A simple example of such a term 
is provided by a Dirac field with a complex spatially (or temporally) 
varying mass term of the form
\be
  \hat m(x) = m_R(x) + i \gamma_5 m_I(x) 
     \equiv \vert m\vert e^{i\gamma^5 \theta}.
\label{mass term}
\ee
This term could be of course replaced by generic scalar and pseudoscalar
operators; we retain the present notation however, because of its clear
physical implications. Namely, this type of term has precisely the form
relevant for baryogenesis, where such an effective mass term
is induced by a coupling of the fermion with 
a spatially varying complex scalar field condensate during a first 
order phase transition.  Here we shall consider a simple toy model with 
the lagrangian
\be
{\cal L}= i \bar \psi \sl \partial \psi - \bar \psi_L \, m \, \psi_R  
                - \bar \psi_R \, m^* \, \psi_L  + {\cal L}_{\rm int} ,
\label{lagr}
\ee
where $m=m_R+im_I$ is a complex mass.
The detailed form of the interaction lagrangian ${\cal L}_{\rm int}$ 
is not relevant for us here. Realistic cases studied in literature
in connection with the CFM baryogenesis include the two 
doublet extension of the standard model \cite{JPT} and the 
chargino sector of the minimal supersymmetric 
standard model \cite{CJK}. 

This letter is a part of a series \cite{JKP1, JKPl, JKP2} dedicated 
to derivation of semiclassical transport equations for quantum fields 
in varying backgrounds beyond the trivial order in gradients. We first 
explicitly construct the fermion propagator for the theory (\ref{lagr}) 
to the first nontrivial order in gradients. Corrections appear as higher 
order poles multiplied by derivatives of the complex phase of the mass. 
The position of the poles is not shifted however, and the physical 
content of the correction terms only becomes apparent after consideration 
of spectral integrals over test functions by a technique 
developed in \cite{JKP1}. It will be shown that the spectral function,
while not expressible as a sum of ordinary delta-functions even in 
the on-shell limit, projects a test function to a sharp, but shifted 
energy shell. We then show that this shell corresponds to the physical 
semiclassical dispersion relation of a standard WKB wave function, and 
describe how the current computed using the spectral function 
representation can be related to motion of WKB wave packets. \\

\section{Propagator}

Our first task is to solve the retarded and advanced propagators 
for the fermionic field $\psi$ in the weak coupling approximation and in an
expansion in gradients of the slowly varying mass terms. We will work
in the Keldysh closed time contour (CTC) formalism \cite{SK,KB,Henning},
and the basic quantity of interest is the path-ordered propagator 
\be
  i G^\psi_{\cal C}(x,y) = \ave{T_{\cal C}[\psi(y)\bar \psi(x)]}.
\label{prop}
\ee
Writing the Schwinger-Dyson equations for $G_{\cal C}$ in the real
time formulation implies the following formally exact equations for
the retarded and advanced propagators $G^{r,a}$ in the Wigner representation
\cite{Henning}
\be
  e^{-i\Diamond } \DBR{(iG_0)^{-1}}{iG^{r,a}(k;X)}=1\,,
\label{propeq1}
\ee
where $\Diamond \DBR{f}{g} \equiv \frac{1}{2}(\partial_X f \cdot
\partial_k g - \partial_k f \partial_X g)$ and the tree level 
propagator is given by
\be
  (iG^{r,a}_0)^{-1} \equiv \sl{k} - m_R(X) - i\gamma_5 m_I(X) 
                   - \Sigma^{r,a} \pm i s_\omega \epsilon,
\label{g0inv}
\ee
where $s_\omega = {\rm sign}(\omega )$, $k$ denotes the canonical 
$4$-momentum and $X \equiv (x+y)/2$ the average position. 
From now on we shall assume a weak coupling limit and neglect the 
self-energy $\Sigma^{r,a} = \Sigma_R \mp i\omega \Gamma$; in 
particular we thus consider the on-shell limit $\Gamma \ra 0$. We 
will also suppress the $i\epsilon$-prescription which differentiates
between $G^{r,a}$ even when $\Gamma \ra 0$, which is introduced as a
mnemonic tool to maintain information about the integration path 
required to fully specify how $G$ acts operationally. A formal 
solution to equation (\ref{g0inv}) can easily be found in the 
gradient approximation. Indeed, to first order in gradients we 
have:
\be
 iG = iG_0 
    - iG_0 \; i\Diamond\DBRb{(iG_0)^{-1}}{iG_0}{(iG_0)^{-1}} \; iG_0,
\label{approx}
\ee
where the free propagator is given by
\be
 iG_0 = \frac{1}{k^2 - \ABS{m}^2}     
     (\sl{k} + m_R(X) - i\gamma_5 m_I(X)).
\label{g0}
\ee
After some algebra one can write the expression (\ref{approx}) into 
the following covariant form:
\be
 i G = i G_0 +  \frac{1}{({k^2 - \ABS{m}^2})^2} 
           \left[i(k\cdot \partial_X - \sl k \sl \partial_X )
                \ABS{m}e^{-i\theta \gamma_5} 
              + \gamma^5 \ABS{m}^2 \sl \partial_X \theta \right]. 
\label{fullG}
\ee
There is clearly a nonvanishing correction to the propagator at the
leading nontrivial order. This is not in disagreement with the no-go
theorem by \cite{Henning}, because, as we shall see, all nontrivial
consequences of (\ref{fullG}) are proportional to the complex 
pseudoscalar mass (or more generally to a pseudoscalar self-energy 
function), which was not considered in Ref.\ \cite{Henning}. An important 
point to observe about (\ref{fullG}) is that the pole of the propagator 
is not shifted, but instead a new second order pole, proportional to 
gradients of $m$ has appeared. This behaviour is analogous to one 
previously observed in the case of a scalar field \cite{JKP1}. 

For simplicity we shall from now on restrict ourselves to the case of a planar
symmetry with only spatially varying mass: $m = m(x_3)$. In this case one can
Lorentz boost to the frame in which ${\bf k}_{\,\parallel}={\bf 0}$, so that
$\omega\rightarrow \tilde\omega 
\equiv s_\omega\nBR{ \omega^2-{\bf k}_{\,\parallel}^{\;2}}^{1/2}$.
With these definitions Eq.~(\ref{fullG}) simplifies to 
\bea
  i \gamma^0 G &=& 
        \frac{1}{z} (\tom - k_3\gamma^5 S^3 
                     + \gamma^0m_R - i\gamma^0\gamma^5m_I)
\nn \\
               &-& \frac{1}{z^2} 
            \BBR{i\tom (  m_R^\prime \gamma^0\gamma^5
                         - i m_I^\prime \gamma^0 )
                         + \ABS{m}^2\theta' }S^3,
\label{planarG}
\eea
where we combined various $\gamma$-matrices into the spin operator 
$S^3 = \gamma^0\gamma^3\gamma^5$, which measures the spin in the 
$\hat {\bf k}_3$-direction and in the frame in which
${\bf k}_\parallel={\bf 0}$, and we used the 
definition $z \equiv k^2 - \ABS{m}^2$. Allowing $z$ a small 
complex value, one sees that (\ref{planarG}) yields $G^r$ 
($G^a$) for ${\rm Im}(z) < 0$ ($ > 0$). A further very important 
simplification arises from the fact that pseudoscalar interactions 
conserve spin (one can show that the spin-operator $S^3$ commutes 
with the propagator $G$). The problem then becomes diagonal 
in spin and, by replacement $S^3\ra s$, the $4\times 4$ structure of 
(\ref{planarG}) becomes block-diagonal, so that the problem 
can be reduced to a $2\times 2$ problem. In the $2\times 2$ 
formalism, it is more convenient to use the Pauli matrices for 
the spinor algebra, which can be effected by the following 
identifications:
\be
          \gamma^0 \; \ra \; \sigma^1, \qquad 
-i\gamma^0\gamma^5 \; \ra \; \sigma^2  \qquad  
         -\gamma^5 \; \ra \; \sigma^3.
\label{subs}
\ee
where we have taken the chiral representation for the $\gamma$ matrices.
In this way (\ref{planarG}) can be recast in a particularly simple 
form:
\bea
i\sigma^1 G_s &=& \frac{1}{z} (\tom + sk_3\sigma^3 
                            + m_R\sigma^1 + m_I\sigma^2)
\nn\\
                &-& \frac{s}{z^2} 
            \BBR{\tom (  m_I^\prime \sigma^1 
                      - m_R^\prime \sigma^2 ) + \ABS{m}^2\theta' }
\nn\\ 
   &\equiv& b_0 + {\bf b}\cdot {\bf \sigma}.
\label{planar2}
\eea
From the last expression in particular it is evident that there exists 
a basis in which the propagator $G$ is diagonal $G_s \ra G_{d,s} = 
b_0 + s\; {\rm sign}(\tom)\ABS{{\bf b}} \sigma^3$. In the Dirac 
notation the propagator then reads
\be
  i \gamma^0 G_d = s_\tom \NBR{\frac{| \tilde \omega |}{z} 
                    - \frac{ss_\tom \ABS{m}^2\theta'}{z^2}} 
                   \NBR{1 - s_\tom S^3\gamma^5},
\label{rotprop}
\ee
where $s_\tom \equiv {\rm sign}(\tom)$.  Physically the equation 
(\ref{rotprop}) means that the states with equal $ss_\tom$ propagate 
identically.  Formally, this is expressed by the projector 
$(1 - s_\tom S^3\gamma^5)/2$, which identifies the states with equal 
entries on the diagonal. In particular this means that particles and 
antiparticles with opposite spin have identical spectral functions. 
However, for particles and antiparticles of {\em same} spin (helicity) 
the spectral functions differ by an amount proportional to the 
CP-violating angle $\theta'$. This is how in the context of baryogenesis
CP-violating backgrounds induce bias in the particle and antiparticle 
distributions. In order to understand how this bias propagates in a plasma, 
it is then necessary to study the relevant quantum trasport equations 
with the effect of collisions included \cite{jkp4}.  

\section{Spectral integrals}

To study the physical consequences of the propagator 
(\ref{fullG}), we now consider spectral integrals over test functions
\cite{JKP1}, representing some generic observables, such as the generalized 
particle distribution function. We define:
\be 
  G[\calT ] \equiv {\rm Tr} \; \frac{i}{2\pi} 
            \int_0^\infty d\omega \gamma^0 \calA \calT ,
\label{Tom1}
\ee
where $\calA\equiv \frac{i}{2} (G^r - G^a)$ is the spectral function,
$\calT$ is a test function, and the trace is taken over the Dirac indices.
Because scalar and pseudoscalar backgrounds considered here conserve spin,
we can take ${\cal T}$ to be diagonal in the spin space and a scalar 
function in a spin $2\times 2$ block. For backgrounds that violate spin 
this analysis would have to be generalized to include spin mixing. 
The $\gamma^0$-factor in Eq.~(\ref{Tom1}) is added to make the projection 
operator explicitly hermitean. The spectral function ${\cal A}$ carries
information on the physical spectrum of excitations in the system. In
particular to zeroth order in gradients it becomes a simple on-shell 
projector
\footnote{Note that we are assuming the on-shell limit and neglecting 
possible thermal and/or vacuum corrections, so that $\Sigma^{r,a} = 0$.}
({\it cf.} Eq.~(\ref{g0})):
\be
{\cal A}\; \ra \;\frac{\sl{k} + m_R - i\gamma_5 m_I}{2\omega_0}
  \left[\delta (\omega - \omega_0) - \delta(\omega + 
\omega_0)\right],
.
\label{A0}
\ee
where $\omega_0 = (k_3^2 + |m|^2)^{1/2}$.  
In a spatially varying background the spectral function can no longer be 
expressed in terms of $\delta$-functions projecting onto the physical 
energy shells.  As discussed in \cite{JKP1}, it can however be written 
as a sum of projectors onto complex shells, but this is not the approach
we will pursue here.
Instead, by extending the momentum to the complex plane, the integral
(\ref{Tom1}) can be turned into a contour integral over the variable 
$z \equiv k^2 - \ABS{m}^2$ \cite{JKP1}:
\be
  G[\calT ] = {\rm Tr}\; \frac{i}{4\pi} \int_{\calC_0} d z 
                \frac{i\gamma^0 G \calT}{\sqrt{\omega_0^2 + z}} ,
\label{Tom2}
\ee
where $G$ is the propagator~(\ref{fullG}), and  $\calC _0$ is a contour
encircling the pole at $z = 0$ ($\omega = \omega_0$) once clockwise.
We can then compute (\ref{Tom2}) by making use of the residue theorem:
\be
  G[\calT ] = {\rm Tr\; Res}_{z=0}
              \BBR{\frac{i\gamma^0 G\calT }{2\sqrt{\omega_0^2 + z}}} .
\label{Tom3}
\ee
Because the highest pole in the expression (\ref{fullG}) is of the
second order, only the terms up to the linear order in $z$ in the expansions 
of various terms appearing in (\ref{Tom3}) contribute. In particular the 
gradient expansion in $\partial_{k_z} \calT$ then terminates at first 
order. That is to say that all $\partial^l_{k_z}\calT$-terms for $l>1$ are of 
at least second order in gradients.  An important consequence
of this truncation is that a complete description of a fermionic plasma, 
consistent to first order in gradients, requires at most two 
independent distribution functions, in contrast to the four distribution
functions required for a full description of the bosonic case at the lowest
nontrivial order in gradients \cite{JKP2}.

It is now a simple matter to compute the residue in Eq.~(\ref{Tom3}) of
the propagator (\ref{planar2}) or (\ref{rotprop}). (For notational simplicity
from now on we shall denote $\tilde \omega$ by $\omega$.) To the order we
are working 
\bea
  G[\calT ] &=& \calT _0 + \frac{ss_\omega\ABS{m}^2\theta'}
             {2\omega_0^2}\nBR{{\calT _0 - \omega_0\calT_0^{\,\prime}}}
\nn\\
            &=& \frac{\omega_0}{\omega_{\rm sc}}\calT_0 (\omega_{\rm sc}),
\label{result1}
\eea
where $\calT_0 \equiv \calT (\omega_0)$, $\calT_0^{\,\prime} \equiv 
(\partial_\omega \calT) (\omega_0)$, and we have defined the semiclassical 
energy as
\be
\omega_{\rm sc} \equiv \omega_0 - \frac{ss_\omega\ABS{m}^2\theta'}
                                    {2\omega_0^2}.
\label{engy1}
\ee
The test function thus still gets projected to a sharply defined energy 
shell, which is however shifted with respect to the lowest order shell
$\omega_0$.

\section{Current}

As a physically motivated application of the above, we now consider a current
of spin states. In the present formalism this can be written as follows:
\be
 j^\mu_s(X) = {\rm Tr} \int \frac{d^4k}{(2\pi )^4} i\gamma^\mu G^<(k;X)P_s, 
\label{curr1}
\ee
where $P_s = (1+sS^3)/2$ is the spin projector and $G^<(k;X)$ is the quantum 
Wigner function.  Assuming the decomposition $iG^<(k;X) P_s = {\cal A} P_s 
n_s$,  where $n_s$ are generalized particle distribution functions of spin 
states, and inserting this expression into (\ref{curr1}) it is a simple 
matter to show that
\bea
 j^0_s(X) = 2 \int \frac{d^3k}{(2\pi )^3}   \frac{\omega_0}{\omega_{\rm sc}}
                   f_s^{\rm sc}({\bf k};X)
\nn \\   
 j^3_s(X) = 2 \int \frac{d^3k}{(2\pi )^3}
           \frac{k_3}{\omega_0} f_s^0({\bf k};X),
\label{j0 j3}
\eea
where the (momentum space) semiclassical distribution function
$f_s^{\rm sc}({\bf k};X)\equiv n_s(\omega_{\rm sc},{\bf k};X)$ is the
projection of $n_s$ onto the semiclassical shell $\omega=\omega_{\rm sc}$,
while the distribution function 
$f_s^0({\bf k};X)\equiv n_s(\omega_0,{\bf k};X)$ is the 
projection of $n_s$ onto the unshifted shell $\omega=\omega_0$. Further,
in Eq.~(\ref{j0 j3}) we have reintroduced the dependencies on the 
trivial perpendicular directions ${\bf k}_\parallel$.
Finally, changing the variables $k_3 \ra k_{\rm sc}$, 
$\omega_{\rm sc}\ra \omega$ 
($dk_3\rightarrow d\omega (\partial\omega_{\rm sc}/dk_3)^{-1}$) 
in the first integral, and
$k_3 \ra k_0\equiv (\omega^2-\vert m\vert^2)^{1/2}$, $\omega_{0}\ra \omega$ 
($dk_3\rightarrow d\omega \omega_{0}/k_3$) 
in the second integral, the current reduces to
\be
 j^\mu_s 
= 2 \int \frac{d\omega d^2k_{||}}{(2\pi )^3 }  \, 
\left(\frac{\omega }{k_{\rm sc}}\,f_s^{\rm sc} ,
  \, \hat {\bf k}f_s^0\right) ,
\label{current}
\ee
where the distribution functions 
$f_s^{\rm sc}=f_s^{\rm sc}(\omega,{\bf k}_\parallel;X)
\equiv n_s(\omega,k_{\rm sc},{\bf k}_\parallel;X)$ and 
$f_s^0 =f_s^0(\omega,{\bf k}_\parallel;X)$
$\equiv n_s(\omega,k_0,{\bf k}_\parallel;X)$ are now defined
on the energy phase space $(\omega,{\bf k}_\parallel;X)$.
The semiclassical momentum 
$k_{\rm sc}=k_{\rm sc}(\omega,{\bf k}_\parallel;X)$ 
in Eq.~(\ref{current}) is the one obtained by inverting 
the semiclassical energy-momentum relation~(\ref{engy1}):
\be
  k_{\rm sc} =  k_0 
        \left(1 + \frac{ss_\omega\ABS{m}^2\theta'}{2k_0\omega^2}\right).
\label{ksc}
\ee
Of course, in equilibrium the spatial currents cancel.  However, if the 
preferred coordinate frame is boosted with respect to the plasma rest frame, 
as is the case for a propagating phase transition front, then there are 
net currents flowing through the space. One can then determine
how these induced currents are affected by the presence of a background 
mass term~(\ref{mass term}). 
Since particles and antiparticles of a given spin (helicity) $s$ and 
energy $\vert\omega\vert$ have different effective momenta $k_{\rm sc}$, 
the corresponding particle and antiparticle densities $j^0_s$ are different 
in the region of a spatially varying background. This is how local 
particle-antiparticle asymmetric density and current profiles get 
created by a spatially varying and CP-violating background, which is 
the phenomenon at the heart of most electroweak baryogenesis mechanisms. 

\section{Field theoretical {\it vs} WKB approach}

Let us now make a connection between the above results and 
the dispersion relation obtained by using the WKB method. 
Through a standard calculation \cite{JPT, CJK}, one finds 
\be
k_{\rm wkb} = k_0 
            + \frac{ss_\omega \theta' }{2k_0}(\omega \pm sk_0) 
            + \alpha' ,
              \qquad k_0 =\sqrt{\omega^2-\ABS{m}^2} ,
\label{WKB dispersion}
\ee
where the arbitrary function $\alpha'$ follows from the reparametrization
invariance of the theory given by the lagrangian (\ref{lagr}) invariant
under (global) $U(1)$-transformations $\psi\rightarrow e^{i\alpha}\psi$. 
It is then obvious that $k_{\rm wkb}$ cannot represent a physical quantity.
The group velocity
\be
  v_g \equiv (\partial_\omega k_{\rm wkb})^{-1},
\label{vqroup}
\ee
corresponding to the stationary phase of the WKB wave packet, is a well
defined physical quantity however. The physical momentum of a WKB state is 
then given by $k_{\rm phys} = \omega v_g(\omega)$. Computing the 
derivative~(\ref{vqroup}) one immediately finds the physical reparametrization 
invariant WKB dispersion relation
\be
  k_{\rm phys} \equiv \omega v_g = k_0 
        \left(1 + \frac{ss_\omega\ABS{m}^2\theta'}{2k_0\omega^2}\right),
\label{kphys}
\ee
which is identical to Eq.~(\ref{ksc}). With this the semiclassical
momentum (\ref{ksc}) can be reinterpreted as the physical momentum of a
WKB state. In this sense we have established the equivalence between the
physical WKB dispersion relation and the field theoretical energy-momentum
shell corresponding to the propagator~(\ref{fullG}), or
equivalently~(\ref{rotprop}). 

 We now return to discuss the current in Eq.~(\ref{current}). The 
difference between the WKB approach and the field theoretical one presented 
here is the fact that there are two mutually independent distribution 
functions $f_s^{\rm sc}$ and $f_s^0$. In the same 
spirit as it was done for the scalar field in Ref. \cite{JKP2}, we may 
introduce a coherent quantum density $f_s^{\rm qc} = f_s^{\rm sc} - f_s^0$,
which is a dynamical measure of quantum coherence on phase space between 
the semiclassical and the unshifted shell. At the leading order in gradients
$f_s^{\rm qc}$ of course vanishes. In analogy to the scalar field case we then 
expect that in the frequent scattering limit $f_s^{\rm qc}$ gets suppressed 
and can be consistently neglected. Assuming this is the case, we may then set
$f_s^{\rm qc} \rightarrow f_s$, $f_s^0 \rightarrow f_s$,
so that finally the current~(\ref{current}) reduces to the following form
\be
 j^\mu_s \rightarrow  2\int \frac{d\omega d^2k_{||}}{(2\pi )^3 }
\left(\frac{1}{v_g}\, ;\, \hat {\bf k}\,\right) f_s\, .
\label{current ii}
\ee
This limit is in perfect agreement with the naive WKB result \cite{ite}, 
implying the following simple physical interpretation: when quasiparticles 
arrive into a region of a spatially varying background, they either
speed up or slow down. If they slow down for example, the local 
particle density increasess proportionally to the factor $1/v_g$. 
At the same time the flux, given by the spatial part of the current,
remains unchanged, as it should.  To prove (\ref{current ii}) rigorously
requires derivation and solution of the relevant quantum transport 
equations to determine selfconsistently the distribution functions 
$f_s^{\rm sc}$ and $f_s^0$. This problem will be further considered in
a forthcoming work \cite{jkp4}.

While it was to be expected that the field theoretical calculation, 
unlike the  WKB-computation, is automatically invariant under 
the field redefinition $\psi\rightarrow e^{i\alpha}\psi$, the fact that 
the two methods lead to identical dispersion relations and currents 
(in the frequent scattering limit) is highly nontrivial. Indeed, it should be
stressed that the physical WKB-shell has emerged from the field theoretical 
calculation indirectly, not as a simple shift of the propagator poles, but
rather by the operation of a modified spectral projector on test functions, 
which is effected {\it via} the technique of spectral integrals.

\section{Conclusions}

We have studied the fermion propagator coupled with a spatially varying
classical background field.  We represented this coupling by a spatially
varying complex mass term, and proved that the propagator receives nontrivial
corrections already at the first order in the gradients of the mass term $m$.
For this result it was crucial that $m$ contains a nonvanishing complex 
pseudoscalar piece. We then explicitly constructed the fermion propagator
and the associated spectral function to the first order in gradients.
The gradient corrections were shown to arise, not as shifts in the poles, 
but rather as additional higher order poles multiplied by the gradients of 
the complex phase of $m$. We then considered a spectral integral of
a simple test function, and showed that it nevertheless gets projected to 
a sharp energy shell, which acquires a spin dependent shift with respect 
to the lowest order shell. The new shell was then shown to coincide with 
the physical dispersion relation of a WKB-state.  We also computed the 
current in the field theoretical way, and showed that it can be 
interpreted in terms of moving WKB wave packets. We note that it is 
straightforward to generalize our results to a generic pseudoscalar 
self-energy term and to temporally varying fields.

Here we have concentrated on the propagation of fermions in a nontrivial
backgrounds. The next logical step is to study the transport properties 
in a plasma containing fermions.  Drawing on the insights obtained from
the study of the scalar case \cite{JKP2}, the results presented here
suggest that a consistent description of transport at nontrivial order
in gradient expansion will require a set of coupled equations for number 
density and an additional function describing quantum coherence in phase 
space.  We also expect that in the frequent scattering limit the quantum
coherence effects are suppressed, and an effective quasiparticle picture 
is recovered \cite{jkp4}.

\section*{Acknowledgements}  
We wish to thank Felipe Freire for collaboration at the early stages of
this project, and Dietrich B\"odecker for illuminating discussions. KK
thanks CERN for hospitality during the completion of this work.

%
%

\nc{\ap}[3]    {{\it Ann.\ Phys.\ }{{\bf #1} {#2} {(#3)}}}
\nc{\ibid}[3]  {{\it ibid.\ }{{\bf #1} {#2} {(#3)}}}
\nc{\jmp}[3]   {{\it J.\ Math.\ Phys.\ }{{\bf #1} {#2} {(#3)}}}
\nc{\np}[3]    {{\it Nucl.\ Phys.\ }{{\bf #1} {#2} {(#3)}}}
\nc{\pl}[3]    {{\it Phys.\ Lett.\ }{{\bf #1} {#2} {(#3)}}}
\nc{\pr}[3]    {{\it Phys.\ Rev.\ }{{\bf #1} {#2} {(#3)}}}
\nc{\prep}[3]  {{\it Phys.\ Rep.\ }{{\bf #1} {#2} {(#3)}}}
\nc{\prl}[3]   {{\it Phys.\ Rev.\ Lett.\ }{{\bf #1} {#2} {(#3)}}}
\nc{\spjetp}[3]{{\it Sov.\ Phys.\ JETP }{{\bf #1} {#2} {(#3)}}}
\nc{\zetp}[3]  {{\it Zh.\ Eksp.\ Teor.\ Fiz.\ }{{\bf #1} {#2} {(#3)}}}

\end{document}